\documentclass[aps,prd,preprintnumbers,superscriptaddress,nofootinbib,notitlepage,floatfix,10pt]{revtex4-2}
\usepackage[pdftex]{graphicx}
\usepackage{bm,latexsym,amsmath,amssymb,amsfonts,mathrsfs}
\usepackage{color}
\allowdisplaybreaks[1]
\usepackage[pdftex,colorlinks=true,linkcolor=blue,citecolor=cyan,backref=page]{hyperref}
\newcommand*{\D}{\mathrm{d}}

\begin{document}
\title{Slowly moving black holes in Lorentz-violating scalar-tensor gravity}
%
\author{Jin~Saito}
\email[Email: ]{j\_saito@rikkyo.ac.jp}
\affiliation{Department of Physics, Rikkyo University, Toshima, Tokyo 171-8501, Japan}
\author{Tsutomu~Kobayashi}
\email[Email: ]{tsutomu@rikkyo.ac.jp}
\affiliation{Department of Physics, Rikkyo University, Toshima, Tokyo 171-8501, Japan}
%
\begin{abstract}
A scalar field with a timelike gradient defines a preferred slicing.
This occurs even in a non-cosmological setup in scalar-tensor theories such as khronometric theory.
We study a black hole moving slowly relative to the preferred slicing defined by the scalar field.
We consider a family of higher-order scalar-tensor theories that extend khronometric theory.
The action is characterized by one constant parameter and one arbitrary function of the kinetic term of the scalar field.
A slowly moving black hole is described by a dipole perturbation of a black hole at rest.
We treat the metric and scalar-field perturbations consistently, derive a single master equation for the coupled system, and reconstruct the metric for a slowly moving black hole from the master variable.
In this way, we revisit and generalize the previous studies.
For generic theories in the family we consider, we find a solution that is regular outside the universal horizon and agnostic to the arbitrary function in the action.
\end{abstract}
\preprint{RUP-24-15}
\maketitle

\section{Introduction}\label{sec:intro}

Scalar fields are ubiquitous in various aspects of gravitational physics and cosmology.
In dark energy models of the late Universe, a scalar field accounts for the accelerated expansion.
The same is of course true in inflationary models of the early Universe.
A number of modified gravity theories beyond general relativity can be described by scalar-tensor theories, which could be an alternative to dark energy or the inflaton.
A scalar field also arises as a low-energy degree of freedom of some fundamental high-energy theory.

Lorentz symmetry is thought of as a fundamental symmetry of nature.
However, if a scalar field $\phi$ has a timelike gradient, $g^{\mu\nu}\partial_\mu\phi\partial_\nu\phi<0$, it defines a preferred slicing and thereby breaks the Lorentz invariance spontaneously.
This is clearly the case in cosmology, but it could also happen in Minkowski spacetime and an asymptotically flat spacetime, depending on the theory.
One of the interesting implications of such a scalar-field configuration is that a moving black hole cannot be obtained by boosting a solution at rest relative to the preferred slicing defined by the scalar field.
A black hole moving through a homogeneous scalar field was recently studied in the case of a massless minimally coupled scalar field in~\cite{Frolov:2023gsl,Frolov:2024htj}.

Motivated by quantum gravity, theories of gravity with broken time-diffeomorphism invariance and hence without full Lorentz invariance have stimulated active discussions~\cite{Liberati:2013xla,Blas:2014aca,Herrero-Valea:2023zex}.
After a seminal work by Ho\v{r}ava~\cite{Horava:2009uw}, different versions of Ho\v{r}ava-Lifshitz gravity have been proposed~\cite{Blas:2009qj,Blas:2010hb,Horava:2010zj}, intending to address the problems present in the original version.
Also in such cases, a scalar field emerges in a fully Lorentz-invariant expression of the action as the St\"{u}ckelberg field of the broken time-diffeomorphism invariance.
The low-energy limit of Ho\v{r}ava-Lifshitz gravity can thus be described by a particular scalar-tensor theory called khronometric theory~\cite{Blas:2009yd,Jacobson:2010mx}.
It is closely related to Einstein-aether theory~\cite{Jacobson:2000xp,Jacobson:2007veq}, in which a preferred frame is determined by a unit timelike vector field.

Moving black holes have been investigated also in khronometric theory.
In~\cite{Ramos:2018oku}, a slowly moving solution was studied by perturbing a black hole at rest, treating the metric and scalar-field perturbations fully consistently.
Reference~\cite{Ramos:2018oku} then found that, for generic values of theory parameters, moving black hole solutions exhibit curvature singularities at the inner boundary called the universal horizon, leaving only a viable one-dimensional parameter space where the moving solution is nothing but Schwarzschild seen in a nontrivial coordinate system.
More recently, Ref.~\cite{Kovachik:2023fos} considered the same problem in the decoupling limit, where the back-reaction of the scalar field on the metric is ignored.
The decoupling limit analysis presented regular scalar-field configurations for generic values of theory parameters, implying a potential contradiction with~\cite{Ramos:2018oku}.

The purpose of the present paper is to revisit and generalize the slowly moving black hole solutions in khronometric theory.
We consider a subset of quadratic higher-order scalar-tensor theories~\cite{Langlois:2015cwa,DeFelice:2018ewo} characterized by one constant parameter and one arbitrary function of the kinetic term of the scalar field.
The subset contains not only khronometric theory, but also a theory with no propagating scalar mode~\cite{Gao:2019twq} which is reminiscent of the cuscuton theory~\cite{Afshordi:2006ad}.
This allows us to draw universal conclusions that can be applied to more general theories than khronometric theory.
We include both metric and scalar-field perturbations consistently as in~\cite{Ramos:2018oku}.
Although the system is composed of a set of complicated coupled differential equations for multiple variables, we can derive a single master equation by solving which one can reconstruct the perturbed metric.
In so doing we separate the part depending on the theory parameters and the theory-independent, universal part in the equations so that one can identify the theory-dependence of the moving black hole solutions.

The rest of this paper is organized as follows.
In the next section, we define a subset of quadratic higher-order scalar-tensor theories with shift symmetry which we shall study.
In Sec.~\ref{sec:BHsoln}, we present a static and spherically symmetric black hole solution at rest with respect to the preferred slicing defined by the scalar field.
In Sec.~\ref{sec:Dipole pert}, we consider static, even-parity, dipole perturbations around this background solution.
We begin by specifying the boundary conditions at infinity so that the perturbed metric describes a moving black hole with respect to the rest frame of the scalar field.
We derive a single master equation and present how all the metric perturbations can be reconstructed from a solution to it.
We then discuss solutions to the master equation, taking care of the regularity of curvature invariants.
Our conclusions are drawn in Sec.~\ref{sec:concluding}.

\section{Quadratic higher-order scalar-tensor theories}\label{sec:Quadratic HOST}

We start from the following action of shift-symmetric higher-order scalar-tensor theories~\cite{Langlois:2015cwa,DeFelice:2018ewo},
\begin{align}
        S=\int\D^4x\sqrt{-g}\left[
                F(X)\mathcal{R}+\sum_{I=1}^5A_I(X)L_I
        \right],\label{eq:HOST-action}
\end{align}
where $\mathcal{R}$ is the Ricci scalar,
\begin{align}
        L_1=\phi_{\mu\nu}\phi^{\mu\nu},\quad L_2=(\Box\phi)^2,
        \quad L_3=\Box\phi\phi^\mu\phi^\nu\phi_{\mu\nu},
        \quad L_4=\phi^\mu\phi_{\mu\rho}\phi^\nu\phi_\nu^\rho,
        \quad 
        L_5=(\phi^\mu\phi^\nu\phi_{\mu\nu})^2,
        \label{eq:Lag1-5}
\end{align}
and $F$ and $A_I$ are functions of $X:=-\phi^\mu\phi_\mu/2$.
Here we use the notations $\phi_\mu=\nabla_\mu\phi$ and $\phi_{\mu\nu}=\nabla_\mu\nabla_\nu\phi$.
We will consider only the case where the gradient of $\phi$ is timelike and hence
we are allowed to take the unitary gauge, $\phi=\phi(t)$.

To characterize scalar-tensor theories described by Eqs.~\eqref{eq:HOST-action} and~\eqref{eq:Lag1-5},
it is convenient to introduce the following combinations of the functions~\cite{Langlois:2017mxy}:
\begin{align}
        &M^2=2(F+2XA_1),
        \quad 
        M^2(1+\alpha_T)=2F,
        \quad 
        M^2(1+\alpha_H)=2(F-2XF_{,X}),
        \notag \\ & 
        M^2\left(1+\frac{2}{3}\alpha_L\right)
        =2(F-2XA_2),
        \quad 
        M^2\beta_1=2\left[
                XF_{,X}-X(A_2-XA_3)
        \right],
        \notag \\ & 
        M^2\beta_2=4X\left[
                A_1+A_2-2X(A_3+A_4)+4X^2A_5
        \right],
        \quad 
        M^2\beta_3=-8\left[
                XF_{,X}+X(A_1-XA_4)
        \right].\label{def:eft-paras}
\end{align}
Some relations are required to be satisfied among these functions to remove the ghost degree of freedom.
Depending on the restrictions on the functions thus imposed, a given theory is classified into Horndeski~\cite{Horndeski:1974wa,Deffayet:2011gz,Kobayashi:2011nu}, degenerate higher-order scalar-tensor (DHOST)~\cite{Langlois:2015cwa,Crisostomi:2016czh}, or unitary degenerate higher-order scalar-tensor (U-DHOST)~\cite{DeFelice:2018ewo} families.
See~\cite{Langlois:2018dxi,Kobayashi:2019hrl} for reviews.

Among a variety of quadratic higher-order scalar-tensor theories,
we are particularly interested in the case where
\begin{align}
        M^2=\text{const.},
        \quad 
        \alpha_T=\alpha_H=\text{const.},
        \quad 
        \beta_1=\beta_2=\beta_3=0,\label{eq:conditions-alpha-beta}
\end{align}
while $\alpha_L$ is a (nonvanishing) free function.
This defines a subset of U-DHOST theories.
In units where $M^2=1$, Eq.~\eqref{eq:conditions-alpha-beta} yields
\begin{align}
        &F=\frac{1}{2}(1+\alpha_T),\quad A_1=-\frac{\alpha_T}{4X},\quad
        A_2=\frac{1}{4X}\left[\alpha_T-\frac{2}{3}\alpha_L(X)\right],
        \notag \\ & 
        A_3=\frac{1}{4X^2}\left[\alpha_T-\frac{2}{3}\alpha_L(X)\right],\quad 
        A_4=-\frac{\alpha_T}{4X^2},\quad A_5=-\frac{\alpha_L(X)}{24X^3}.
        \label{eq:def-functions}
\end{align}
The theory is thus characterized by a constant parameter $\alpha_T$
and a function $\alpha_L(X)$.

Let us motivate the study of the above theory.
Recently, the post-Newtonian limit of quadratic higher-order scalar-tensor theories
with $\alpha_L\neq 0$ was derived in~\cite{Saito:2024xdx}.
In the theory defined by the functions~\eqref{eq:def-functions},
the parametrized post-Newtonian (PPN) parameters~\cite{Will:2018bme} are given by~\cite{Saito:2024xdx}
\begin{align}
    &\gamma^{\textrm{PPN}}=1,\quad 
    \beta^{\textrm{PPN}}=1,\quad 
    \alpha_1^{\textrm{PPN}}=8\alpha_T,
    \quad 
    \alpha_2^{\textrm{PPN}}
    =\alpha_T+\frac{3\alpha_T^2}{1+\alpha_T}\cdot 
    \frac{1+\alpha_L}{\alpha_L},
    \notag \\ 
    & \xi^{\textrm{PPN}}=\alpha_3^{\textrm{PPN}}=
    \zeta_1^{\textrm{PPN}}=\zeta_2^{\textrm{PPN}}=\zeta_3^{\textrm{PPN}}=\zeta_4^{\textrm{PPN}}=0,
\end{align}
while the propagation speed of gravitational waves, $c_{\textrm{GW}}$,
is related to $\alpha_T$ as 
\begin{align}
        c_{\textrm{GW}}^2=1+\alpha_T.
\end{align}
Setting $\alpha_T=0$, all the PPN parameters coincide with the values in general relativity irrespective of $\alpha_L$.
Therefore, this is the physically most interesting case.
Nevertheless, in this paper, we treat $\alpha_T$ as a free parameter
because introducing nonvanishing $\alpha_T$ does not change our analysis essentially
and is convenient for comparing our results with previous ones.

In terms of the unit timelike vector defined as $u^\mu=-\phi^\mu/\sqrt{2X}$,
the action can be written as 
\begin{align}
        S=\frac{1+\alpha_T}{2}\int\D^4x\sqrt{-g}\left[
                \mathcal{R}
                +\frac{\alpha_T-2\alpha_L/3}{1+\alpha_T}(\nabla_\mu u^\mu)^2 
                -\frac{\alpha_T}{1+\alpha_T} \nabla_\mu u_\nu \nabla^\nu u^\mu
        \right].\label{eq:vector-action}
\end{align}
This expression reminds us of the action for Einstein-aether theory~\cite{Jacobson:2000xp} and
khronometric theory~\cite{Blas:2009yd,Jacobson:2010mx}, though in the present case $\alpha_L$ is no longer a constant parameter
but is promoted to a function of $X$.

The action takes a simpler form in the unitary gauge: 
\begin{align}
        S=\frac{1}{2}\int\D t\D^3x\sqrt{\gamma}N 
        \left[
                K_{ij}K^{ij}-\left(1+\frac{2}{3}\alpha_L\right)K^2+(1+\alpha_T)R 
        \right],\label{eq:ADM-action}
\end{align}
where $K_{ij}$ is the extrinsic curvature and $R$ is the three-dimensional Ricci scalar of
constant time hypersufraces.
In the unitary gauge, $\alpha_L$ is regarded as a function of the lapse function $N$.


We now provide two specific examples within the above class of scalar-tensor theories.
The first one is khronometic theory~\cite{Blas:2009yd,Jacobson:2010mx} and the infrared limit of Ho\v{r}ava-Lifshitz gravity~\cite{Horava:2009uw,Blas:2009qj},
in which
\begin{align}
    \alpha_L=\mathrm{const.}
\end{align}
However, this represents only a subset of the whole parameter space of khronometric theory.
To cover the entire parameter space, we need to introduce another constant parameter corresponding to $\beta_3$ in Eq.~\eqref{def:eft-paras}.
The present case, $\beta_3\to 0$, amounts to considering the limit of the infinite propagation speed of the scalar mode (see Appendix.~\ref{sec:app-pert-Mink}).
Note that Khronometric theory with $\beta_3=0$ has odd dimensionality of the phase space at each spacetime point~\cite{Li:2009bg,Henneaux:2009zb}.

The second example is the theory with two tensor degrees of freedom and no propagating scalar degree of freedom~\cite{Gao:2019twq}.
In this case, $\alpha_L(X)$ is given by
        \begin{align}
                \alpha_L=-\frac{b_L\sqrt{2X}}{b_L\sqrt{2X}+1}
                =-\frac{b_L}{b_L+N},\label{eq:2dof-alpha-L}
        \end{align}
with $b_L$ being a constant parameter of the theory.
This example has the four-dimensional phase space at each spacetime point~\cite{Gao:2019twq}.
The theory is reminiscent of the cuscuton theory~\cite{Afshordi:2006ad} and its extensions~\cite{Iyonaga:2018vnu}.


\section{Black hole at rest with respect to the preferred slicing}\label{sec:BHsoln}

Let us present a static and spherically symmetric metric with
a time-dependent scalar field that defines the preferred slicing.
It is straightforward to show that the theory we are considering admits
the following solution for generic $\alpha_L(X)$:
\begin{align}
        \phi&=\frac{t}{N_0},
        \\ 
        g_{\mu\nu}\D x^\mu\D x^\nu&=-\left(f-B^2\right)\D t^2+\frac{2B}{\sqrt{f}}\D t\D r
         + \frac{\D r^2}{f}+r^2\D\Omega^2,
\end{align}
where $\D\Omega^2=\D\theta^2+\sin^2\theta\D\varphi^2$ and 
\begin{align}
        f=1-\frac{\mu_0}{r}+\frac{b_0^2}{r^4},
        \quad 
        B=\frac{(1+\alpha_T)^{1/2}b_0}{r^2}.
\end{align}
The solution has three integration constants, $\mu_0$, $N_0$, and $b_0$.
The corresponding Arnowitt-Deser-Misner (ADM) variables are given by 
\begin{align}
        N=\sqrt{f(r)},
        \quad 
        N_i\D x^i=\frac{B(r)}{\sqrt{f(r)}}\D r,
        \quad 
        \gamma_{ij}\D x^i\D x^j=\frac{\D r^2}{f(r)}+r^2\D\Omega^2.
\end{align}
One can see that the trace of the extrinsic curvature vanishes: $K=-\partial_r(r^2B)/r^2=0$.
For this reason, this solution is obtained irrespective of $\alpha_L$.
The solution is essentially the same as the one obtained in Einstein-aether theory~\cite{Berglund:2012bu}.
In Appendix~\ref{app:mono}, we study static monopole perturbations of the above configuration and show that there are no nontrivial solutions other than those corresponding to the shifts of the integration constants $\mu_0$ and $b_0$.

To see the geometrical structure of the solution more clearly,
let us use the new time coordinate $T$ defined by 
\begin{align}
        \D T=\D t -\frac{B\D r}{(f-B^2)\sqrt{f}},
\end{align}
and write the metric as 
\begin{align}
        g_{\mu\nu}\D x^\mu\D x^\nu = 
        -h\D T^2+\frac{\D r^2}{h}+r^2\D\Omega^2,
\end{align}
where 
\begin{align}
        h:=f-B^2=1-\frac{\mu_0}{r^2}-\frac{\alpha_Tb_0^2}{r^4}.
\end{align}
This shows that for $\alpha_T=0$ the geometry is nothing but Schwarzschild.

The location derived from the condition $N=\sqrt{f(r)}=0$ is called the universal horizon~\cite{Barausse:2011pu,Blas:2011ni}.
For the universal horizon to be regular, we require that 
\begin{align}
        b_0=\frac{3\sqrt{3}\mu_0^2}{16}.
\end{align}
For this value of $b_0$ we have 
\begin{align}
        f=\frac{(4r-3\mu_0)^2(16r^2+8r\mu_0+3\mu_0^2)}{256r^4}\,(\ge 0).
\end{align}
Without loss of generality, we set $\mu_0=1$ and accordingly $b_0=3\sqrt{3}/16$ in the following analysis.
The location of the universal horizon is then given by $r=3/4$.
Now we see that the solution describes a black hole.
In particular, it is at rest with respect to the preferred slicing defined by the scalar field.

\section{Dipole perturbations and slowly moving black holes}\label{sec:Dipole pert}

Let us study a black hole moving slowly (along the $z$-axis) with respect to the preferred slicing.
To do so, we consider even-parity dipole perturbations of the spherically symmetric solution presented
in the previous section.
We will not treat the scalar-field perturbation as a test field.
Rather, we will perturb both the scalar field and the metric, and
by the use of the gauge degree of freedom we will set the scalar-field perturbation to zero,
i.e. we will work in the unitary gauge.
It is found that the system of the field equations can be reduced to a single fourth-order master equation, which allows us to reconstruct all the components of the metric perturbations in the unitary gauge from a solution to the master equation.
The following analysis thus advances the previous study~\cite{Kovachik:2023fos} by taking into account the perturbations of the scalar field and the metric consistently, though we cover only the cases with the infinite propagation speed of the scalar mode.
Our master equation and the conclusions derived from it can be applied not only to constant $\alpha_L$, but also to general $\alpha_L(X)$.
In this respect, we generalize and revisit the analysis of~\cite{Ramos:2018oku}.

\subsection{Master equation for dipole perturbations}

The even-parity, stationary dipole perturbations
that can be used to describe a black hole moving along the $z$-axis are given by
\begin{align}
        \phi&=\frac{t}{N_0}+\delta\phi\cos\theta,
        \\ 
        \delta g_{\mu\nu}\D x^\mu\D x^\nu &= 
        -\mathsf{H}_0\cos\theta \D t^2 + \frac{2\mathsf{H}_1}{\sqrt{f}}\cos\theta \D t\D r
        -2\mathsf{b}\sin\theta\D t\D \theta 
        +\frac{\mathsf{H}_2}{f}\cos\theta \D r^2-2\mathsf{a}\sin\theta \D r\D\theta 
        +r^2\mathsf{K}\cos\theta\D\Omega^2,
\end{align}
where $\delta\phi$, $\mathsf{H}_0$, $\mathsf{H}_1$, $\mathsf{H}_2$, $\mathsf{a}$, $\mathsf{b}$,
and $\mathsf{K}$ are small perturbations that depend only on $r$.
Under an infinitesimal coordinate transformation 
\begin{align}
        \bar t=t+\xi_t(r)\cos\theta,\quad \bar r=r+\xi_r(r)\cos\theta,\quad \bar\theta=\theta-\xi_\Omega(r)\sin\theta,
\end{align}
the perturbations transform as 
\begin{align}
        &\overline{\delta\phi}=\delta\phi-\xi_t,
        \quad 
        \bar{\mathsf{H}}_0=\mathsf{H}_0-(f-B^2)'\xi_r,\quad
        \bar{\mathsf{H}}_1=\mathsf{H}_1+\sqrt{f}(f-B^2)\xi_t'
        -B\xi_r'-\sqrt{f}\left(\frac{B}{\sqrt{f}}\right)'\xi_r,
        \notag \\ & 
        \bar{\mathsf{H}}_2=\mathsf{H}_2-2\sqrt{f}B\xi_t'-2\xi_r'+\frac{f'}{f}\xi_r,
        \quad 
        \bar{\mathsf{a}}=\mathsf{a}-\frac{\xi_r}{f}-r^2\xi_\Omega'-\frac{B}{\sqrt{f}}\xi_t,
        \quad 
        \bar{\mathsf{b}}=\mathsf{b}-\frac{B}{\sqrt{f}}\xi_r+(f-B^2)\xi_t,
        \notag \\ & 
        \bar{\mathsf{K}}=\mathsf{K}-\frac{2\xi_r}{r}+2\xi_\Omega,
        \label{eq:gauge-tr-rules}
\end{align}
where a prime denotes differentiation with respect to $r$.
From this, we see that 
\begin{align}
        \bar{\mathsf{H}}_0+2B\bar{\mathsf{H}}_1-B^2\bar{\mathsf{H}}_2=
        \mathsf{H}_0+2B\mathsf{H}_1-B^2\mathsf{H}_2+2Bf^{3/2}\xi_t'-f'\xi_r.
\end{align}
Using the above gauge degrees of freedom, we choose to set
\begin{align}
        \delta\phi=0,\quad \mathsf{H}_0=-2B\mathsf{H}_1+B^2\mathsf{H}_2,\quad \mathsf{K}=0,
\end{align}
which is the same gauge conditions used in~\cite{Saito:2023bhn}.
In this coordinate system, the scalar field is at rest.

To impose appropriate boundary conditions in the unitary gauge,
it is convenient to move temporarily to the frame where the black hole is at rest.
In the rest frame of the black hole, the scalar field at $r\to \infty$ must be of the form~\cite{Kovachik:2023fos} 
\begin{align}
        \phi\approx \frac{t}{N_0}+\mathtt{v}r\cos\theta,
\end{align}
where $\mathtt{v}$ is the (constant) velocity of the black hole
with respect to the rest frame of the scalar field.
Thus, denoting with a bar the perturbations in the rest frame of the black hole,
we have
\begin{align}
        \overline{\delta\phi}\approx \mathtt{v}r,
        \quad \overline{\delta g}_{\mu\nu}\approx 0,
\end{align}
at $r\to \infty$.
This coordinate system can be obtained by applying the gauge transformation
$\xi_t=-\mathtt{v}r$ to the unbarred frame where the scalar field is at rest.
It then follows from the gauge transformation rules~\eqref{eq:gauge-tr-rules} that
\begin{align}
        \mathsf{H}_1\approx \mathtt{v},
        \quad \mathsf{H}_2\approx 0, \quad \mathsf{a}\approx 0,
        \quad \mathsf{b}\approx \mathtt{v}r,
        \label{eq:bc-inf}
\end{align}
for large $r$ in the unbarred frame.
These are the boundary conditions we impose on the unbarred quantities.


We substitute the perturbed metric to the action and expand it to second order in perturbations.
The resultant quadratic action for the static perturbations is given by 
\begin{align}
        S=4\pi \int\D t\D r\mathcal{L}^{(2)},
\end{align}
with 
\begin{align}
        \mathcal{L}^{(2)}&=-\frac{r^2\alpha_L}{9}
        \left[
                \mathsf{H}_1'+\frac{2}{r}\mathsf{H}_1-\frac{B}{2}\mathsf{H}_2' 
                +\frac{3}{\alpha_L r}\left(\mathsf{H}_1-B\mathsf{H}_2\right)
                +\frac{1}{r^2}\left(2+\frac{3}{\alpha_L}\right)\left(B\mathsf{a}-\frac{\mathsf{b}}{\sqrt{f}}\right)
        \right]^2
        \notag \\ & \quad 
        +\frac{1}{6}\left[
                \mathsf{b}'-\frac{2}{r}\mathsf{b}+\frac{1}{\sqrt{f}}\left(
                        \mathsf{H}_1-B\mathsf{H}_2 
                \right)
        \right]^2+\frac{1+\alpha_L}{\alpha_L}\left[
                \mathsf{H}_1-B\mathsf{H}_2+\frac{1}{r}\left(
                        B\mathsf{a}-\frac{\mathsf{b}}{\sqrt{f}}
                \right)
        \right]^2
        \notag \\ & \quad 
        +\frac{1}{12r^2}\left\{
        4(1+\alpha_T)f\mathsf{a}^2+2r\left[
                12B\mathsf{H}_1+(1+\alpha_T)(rf'-2)\mathsf{H}_2
        \right]\mathsf{a}-r\left[
                \frac{12B\mathsf{b}}{\sqrt{f}}-r(1+\alpha_T)(rf)'\mathsf{H}_2
        \right]\mathsf{H}_2
        \right\}.\label{eq:Lag-pert-01}
\end{align}
Here, $\alpha_L$ is evaluated at the background (i.e. $X=[2N_0^2f(r)]^{-1}$),
and hence is dependent on $r$ in general.

By rescaling the variables as 
\begin{align}
        \mathsf{H}_{2\mathrm{new}}=(1+\alpha_T)^{1/2}\mathsf{H}_2,
        \quad \mathsf{a}_{\mathrm{new}}=(1+\alpha_T)^{1/2}\mathsf{a},
\end{align}
the above Lagrangian reduces to the one for $\alpha_T=0$.
(Note that $B\propto(1+\alpha_T)^{1/2}$.)
This property implies that the $\alpha_T$-dependence of a moving black hole solution must be rather trivial.

We closely follow the procedure in~\cite{Saito:2023bhn} to derive the quadratic Lagrangian written solely in terms of a single master variable.
Introducing auxiliary variables $\mathsf{P}(r)$ and $\mathsf{Q}(r)$,
the Lagrangian~\eqref{eq:Lag-pert-01} can be written equivalently as 
\begin{align}
        \mathcal{L}^{(2)}_{\mathsf{PQ}}&=\mathcal{L}^{(2)}
        +\frac{r^2\alpha_L}{9}
        \left[
                \mathsf{H}_1'+\frac{2}{r}\mathsf{H}_1-\frac{B}{2}\mathsf{H}_2' 
                +\frac{3}{\alpha_L r}\left(\mathsf{H}_1-B\mathsf{H}_2\right)
                +\frac{1}{r^2}\left(2+\frac{3}{\alpha_L}\right)\left(B\mathsf{a}-\frac{\mathsf{b}}{\sqrt{f}}\right)
                -\frac{\mathsf{Q}}{\alpha_L}
        \right]^2
        \notag \\ &\quad 
        -\frac{1}{6}\left[
                \mathsf{b}'-\frac{2}{r}\mathsf{b}+\frac{1}{\sqrt{f}}\left(
                        \mathsf{H}_1-B\mathsf{H}_2 
                \right)-\mathsf{P}
        \right]^2.
\end{align}
Indeed, by substituting the solutions to the equations of motion for $\mathsf{P}$ and $\mathsf{Q}$,
one finds that the additional terms vanish and $\mathcal{L}^{(2)}_{\mathsf{PQ}}$ reduces to $\mathcal{L}^{(2)}$.
As one will encounter quite messy expressions at each intermediate step below,
we will not present them explicitly.
Performing integration by parts in $\mathcal{L}^{(2)}_{\mathsf{PQ}}$, we can remove all the derivatives acting on $\mathsf{H}_1$, $\mathsf{H}_2$, $\mathsf{a}$, and $\mathsf{b}$.
Thus, the equations of motion for these variables are constraint equations, enabling us to express $\mathsf{H}_1$, $\mathsf{H}_2$, $\mathsf{a}$, and $\mathsf{b}$ in terms of $\mathsf{P}$, $\mathsf{Q}$, and their derivatives.
Substituting the results back to $\mathcal{L}^{(2)}_{\mathsf{PQ}}$, we obtain a quadratic Lagrangian written in terms of $\mathsf{P}$ and $\mathsf{Q}$.
Now, we replace $\mathsf{Q}$ with the new variable $\chi$ defined by
\begin{align}
        \mathsf{Q}=\sqrt{f}\left(\frac{\chi}{r^3}+\frac{3}{2r}\mathsf{P}\right).
\end{align}
In the resultant Lagrangian, there is no derivative acting on $\mathsf{P}$.
We can thus remove $\mathsf{P}$ by substituting the solution to the equation of motion for $\mathsf{P}$.
We finally arrive at
\begin{align}
        \mathcal{L}_\chi=\frac{4}{3^6}\left[d_1(r)\left(\chi''\right)^2 
        +d_2(r)\left(\chi'\right)^2+d_3(r)\chi^2\right],
        \label{eq:Lag-chi}
\end{align}
where 
\begin{align}
        d_1&=A_Lr^2f^3,
        \\ 
        d_2&=\frac{1}{2}A_L f^2\left(12f +8r f'-r^2f''\right)-\frac{1}{2}
        \left(A_Lf^{1/2}\right)'rf^{3/2}\left(4f+3rf'\right),
        \\ 
        d_3&=A_Lr^{-1}f\left(12f'+rf''\right)
        +\left(A_Lf^{1/2}\right)'r^{-1}f^{1/2}\left(4f-3rf'\right)
        -2\left(A_Lf^{1/2}\right)''f^{3/2},
\end{align}
with 
\begin{align}
    A_L(r):=-\frac{\alpha_L}{1+\alpha_L}.
\end{align}

The fourth-order equation of motion derived from the Lagrangian~\eqref{eq:Lag-chi} can be written as 
\begin{align}
        \zeta''-V(r)\zeta=0,\label{eq:eom-zeta}
\end{align}
where 
\begin{align}
        \zeta(r) &= A_L r f^{9/4}\left[\chi''+\frac{\left(r^2f^{3/2}\right)'}{r^2f^{3/2}}\chi'-\frac{2}{r^2f}\chi\right],\label{eq:def-zeta}
        \\ 
        V(r)&=\frac{2}{r^2f}+\frac{\left(rf^{3/4}\right)''}{rf^{3/4}}.
        \label{eq:def-V}
\end{align}
Remarkably, in terms of
\begin{align}
        \psi:= rf^{3/4}\chi,
\end{align}
Eq.~\eqref{eq:def-zeta} is written as
\begin{align}
        \zeta=A_L f^{3/2}\left(\psi''-V\psi\right),\label{eq:psi-eom}
\end{align}
where one notices the same structure as that of Eq.~\eqref{eq:eom-zeta} in the right-hand side.
It should be noted that the master equation is independent of $\alpha_T$.
Furthermore, the $\alpha_L$-dependence appears only in the overall factor in Eq.~\eqref{eq:def-zeta}.

In terms of $\chi$ and $\zeta$, one can express the metric perturbations in the unitary gauge as 
\begin{align}
        \mathsf{H}_1&=
        \eta_{\mathsf{H}_1}^{(1)}(r)\chi +\eta_{\mathsf{H}_1}^{(2)}(r)\chi' 
        +\left[\eta_{\mathsf{H}_1}^{(3)}(r)+\frac{\tilde\eta_{\mathsf{H}_1}^{(3)}(r)}{A_L}\right]
        \zeta+\eta_{\mathsf{H}_1}^{(4)}(r)\zeta',
        \label{eq:eta-1}
        \\ 
        \sqrt{3}(1+\alpha_T)^{1/2}\mathsf{H}_2
        &=\eta_{\mathsf{H}_2}^{(1)}(r)\chi +\eta_{\mathsf{H}_2}^{(2)}(r)\chi' 
        +\left[\eta_{\mathsf{H}_2}^{(3)}(r)+\frac{\tilde\eta_{\mathsf{H}_2}^{(3)}(r)}{A_L}\right]
        \zeta+\eta_{\mathsf{H}_2}^{(4)}(r)\zeta',
        \label{eq:eta-2}
        \\ 
        \sqrt{3}(1+\alpha_T)^{1/2}\mathsf{a}
        &=\eta_{\mathsf{a}}^{(1)}(r)\chi +\eta_{\mathsf{a}}^{(2)}(r)\chi' 
        +\left[\eta_{\mathsf{a}}^{(3)}(r)+\frac{\tilde\eta_{\mathsf{a}}^{(3)}(r)}{A_L}\right]
        \zeta+\eta_{\mathsf{a}}^{(4)}(r)\zeta',
        \label{eq:eta-3}
        \\ 
        \mathsf{b}&=\eta_{\mathsf{b}}^{(1)}(r)\chi +\eta_{\mathsf{b}}^{(2)}(r)\chi' 
        +\eta_{\mathsf{b}}^{(3)}(r)
        \zeta+\eta_{\mathsf{b}}^{(4)}(r)\zeta',
        \label{eq:eta-4}
\end{align}
where the coefficients $\eta_{\mathsf{H}_1}^{(1)}, \dots$ are
universal functions of $r$ in the sense that they do not depend on $\alpha_T$ and $A_L$.
The explicit expressions for these coefficients are presented in Appendix~\ref{sec:app-eta-cfs}.
As already mentioned above, the $\alpha_T$-dependence of the metric perturbations appears only in the overall factor in Eqs.~\eqref{eq:eta-2} and~\eqref{eq:eta-3}.
Thus, our master equation and the metric reconstruction allow us to see clearly the parameter dependence of a moving black hole solution.

It is interesting to compare the master equation we have just derived with the result in the decoupling limit obtained by taking only the scalar-field perturbation into account and ignoring the metric perturbations.
The decoupling limit has often been utilized in the literature~\cite{Lin:2017jvc,Kovachik:2023fos}.
We substitute the fixed background metric presented in Sec.~\ref{sec:BHsoln} and
\begin{align}
    \phi=\frac{t}{N_0}+\pi (r)\cos\theta 
\end{align}
to the scalar-field equation, $\delta S/\delta\phi=0$.
To linear order in $\pi(r)$, the field equation reads
\begin{align}
    \tilde\zeta''-V(r)\tilde\zeta
    +\frac{1}{6}\left(\frac{3}{2r}\right)^7\alpha_T^2 f^{1/4}\pi = 0,
    \label{eq:testf-1}
\end{align}
where $V(r)$ here is the quantity defined in Eq.~\eqref{eq:def-V} and 
\begin{align}
    \tilde\zeta = \alpha_L rf^{9/4}
    \left[
    \pi'' +\frac{\left(r^2f^{3/2}\right)'}{r^2f^{3/2}}\pi'-\frac{2}{r^2f}\pi 
    \right],
    \label{eq:testf-2}
\end{align}
with $\alpha_L=\alpha_L|_{X=(2N_0^2f)^{-1}}$.
We have thus obtained almost the same equation as the master equation.
The differences are the term proportional to $\alpha_T^2$ in Eq.~\eqref{eq:testf-1} and the prefactor $\alpha_L$ in Eq.~\eqref{eq:testf-2}.
To linear order in $\alpha_T$ and $\alpha_L$, the scalar-field equation in the decoupling limit coincides with the master equation.
It should be emphasized, however, that the master equation has been derived by treating all the perturbations consistently without assuming that $\alpha_T$ and $\alpha_L$ are small.

\subsection{Solution to the master equation}

Let us first show that $\zeta=0$ is required from the regularity of curvature invariants at the universal horizon.
One can see from Eq.~\eqref{eq:def-V} that $V(r)$ is singular only at the universal horizon, $r=3/4$.
Near $r=3/4$, one has 
\begin{align}
    V=\frac{7}{4\varepsilon^2}+\mathcal{O}(\varepsilon^{-1}),
\end{align}
with $\varepsilon:=r-3/4$.
The general solution to Eq.~\eqref{eq:eom-zeta} in the vicinity of the universal horizon is therefore given by
\begin{align}
        \zeta \simeq \mathcal{Z}_+\varepsilon^{1/2+\sqrt{2}}+\mathcal{Z}_-\varepsilon^{1/2-\sqrt{2}},
\end{align}
where $\mathcal{Z}_+$ and $\mathcal{Z}_-$ are integration constants.
To derive the behavior of $\chi$ near the universal horizon, one must solve Eq.~\eqref{eq:def-zeta} with $\zeta$ given above.
However, this step is not necessary in excluding $\zeta\neq 0$.

We now move to check the regularity of curvature invariants at the universal horizon.
To avoid gauge ambiguities, we look at the quantities that vanish when evaluated at the background metric.\footnote{We thank Masashi Kimura for pointing this out.}
For $\alpha_T=0$, $\mathcal{R}$ is one of the appropriate quantities.
When $\alpha_T\neq 0$, we have $\mathcal{R}\neq 0$ for the background metric because the geometry is no longer Schwarzschild. However, it is easy to see that the combination
\begin{align}
    \mathcal{I}:=\mathcal{R}_{\mu\nu}\mathcal{R}^{\mu\nu}-\frac{5}{2}\mathcal{R}^2
\end{align}
vanish for the background metric irrespective of $\alpha_T$.
We therefore evaluate $\mathcal{R}$ in the case of $\alpha_T=0$ and $\mathcal{I}$ in the case of $\alpha_T\neq 0$ for the perturbed metric.
Without using the explicit expressions for the solutions for $\zeta$ and $\chi$,
one can write the curvature invariants in terms of $\zeta$, $\chi$, and their derivatives by the use of Eqs.~\eqref{eq:eta-1}--\eqref{eq:eta-4}.
Using then Eqs.~\eqref{eq:eom-zeta} and~\eqref{eq:def-zeta}, one can remove second and higher derivatives of $\zeta$ and $\chi$ to express the curvature invariants in terms of $\zeta$, $\chi$, and their first derivatives.
In the case of $\alpha_T=0$, $\mathcal{R}$ turns out to be independent of $\chi$ and is given by
\begin{align}
    \frac{\mathcal{R}}{\cos\theta}=-\frac{\sqrt{3}}{12}\frac{1}{r^2}
    \left(\frac{\zeta}{rf^{3/4}}\right)'.
\end{align}
Since $f=\mathcal{O}(\varepsilon^2)$, $\mathcal{R}$ entails terms proportional to $\mathcal{O}(\mathcal{Z}_\pm \varepsilon^{-2\pm\sqrt{2}})$, which are singular at $\varepsilon=0$.
Therefore, $\zeta=0$ is required from the regularity of $\mathcal{R}$.
In the case of $\alpha_T\neq 0$, we rather look at $\mathcal{I}$, which is given by 
\begin{align}
    \frac{256}{27}\left(\frac{1+\alpha_T}{3}\right)^{1/2}\cdot\frac{\mathcal{I}}{\cos\theta}
    &=-\frac{3\alpha_T^2}{2r^{12}}\left(r^2f^{1/2}\chi\right)'+
    \frac{1}{r^8}\left\{\left[
    2\alpha_T(1+\alpha_T)
    +\frac{3\alpha_T^2}{2A_L}
    \right]\frac{\zeta}{rf^{3/4}}\right\}'
    \notag \\ &\quad 
    -\alpha_T^2\left[
    \frac{f}{r^7}\left(\frac{\zeta}{rf^{3/4}}\right)'-\frac{\zeta}{r^9f^{3/4}}
    \right].
\end{align}
Also in this case, $\mathcal{I}$ diverges as $\sim \mathcal{Z}_\pm \varepsilon^{-2\pm\sqrt{2}}$ at the universal horizon unless $\zeta=0$.
Thus, $\zeta\neq 0$ is excluded.
Note in passing that $\mathcal{I}$ vanishes when $\alpha_T=0$.

It is therefore sufficient to consider Eq.~\eqref{eq:def-zeta} with $\zeta =0$, which indicates that the solution is agnostic to $\alpha_L(X)$.
The general solution in the vicinity of the universal horizon is given by
\begin{align}
    \chi\simeq \mathcal{C}_+\varepsilon^{-1+\sqrt{2}}+\mathcal{C}_-\varepsilon^{-1-\sqrt{2}},
    \label{eq:soln-chi}
\end{align}
where $\mathcal{C}_+$ and $\mathcal{C}_-$ are integration constants.
We now have $\zeta=0$, and hence $\mathcal{R}=0$ even though $\chi$ diverges if $\mathcal{C}_-\neq 0$.
In the case of $\alpha_T=0$, the solution~\eqref{eq:soln-chi} is therefore allowed.
On the other hand, $\mathcal{I}\sim \mathcal{C}_\pm \varepsilon^{-1\pm\sqrt{2}}$ and hence we set $\mathcal{C}_-=0$ to avoid the divergence of $\mathcal{I}$ in the case of $\alpha_T\neq 0$.
Note, however, that the resultant solution still has non-analyticity at the universal horizon in the sense that the first derivative of $\mathcal{I}$ diverges there.
Having derived the behavior of $\chi$ in the vicinity of $r=3/4$,
one can integrate Eq.~\eqref{eq:def-zeta} (with $\zeta=0$) from the universal horizon outward, and then reconstruct the metric perturbations using Eqs.~\eqref{eq:eta-1}--\eqref{eq:eta-4}.
This step can be done numerically without specifying $\alpha_T$ and $\alpha_L(X)$.
The reconstructed metric perturbations are independent of $\alpha_L(X)$, and the $\alpha_T$-dependence appears only in a trivial way through Eqs.~\eqref{eq:eta-2} and~\eqref{eq:eta-3}.

\begin{figure}[tb]
  \begin{minipage}[b]{0.48\linewidth}
    \centering
    \includegraphics[keepaspectratio=true,height=55mm]{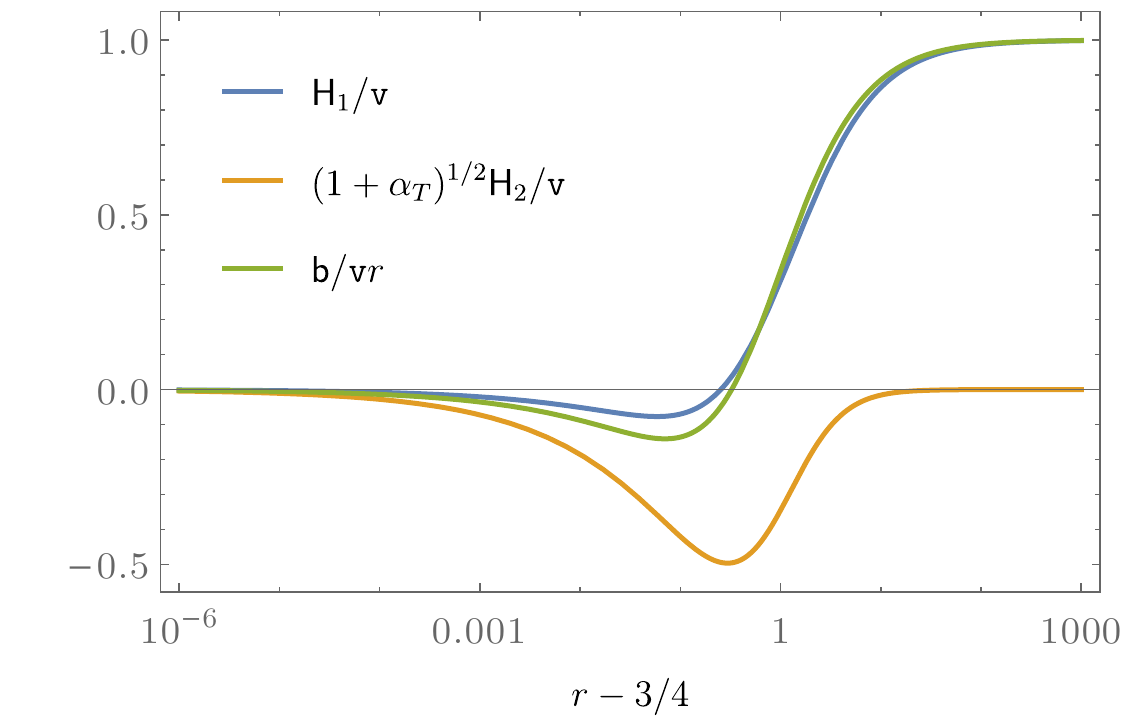}
  \end{minipage}
  \begin{minipage}[b]{0.48\linewidth}
    \centering
    \includegraphics[keepaspectratio=true,height=55mm]{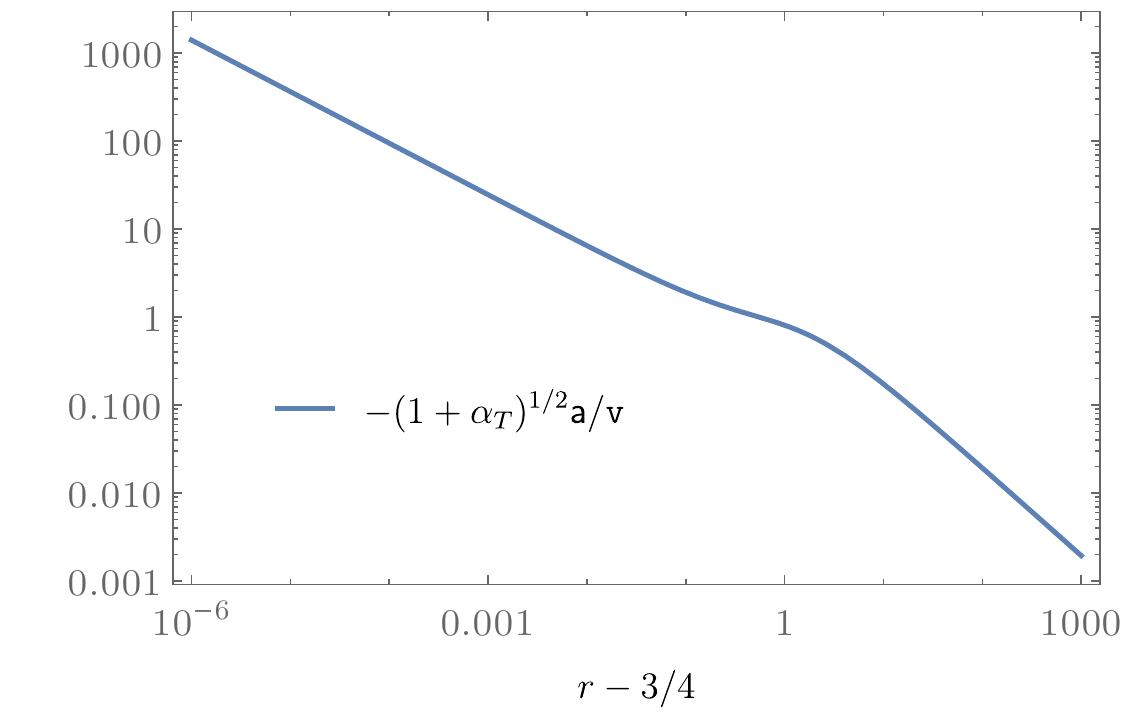}
  \end{minipage}
  \caption{Metric perturbations reconstructed from the master variable as functions of $r$.
  The $\alpha_T$ dependence is isolated as the overall factor $(1+\alpha_T)^{1/2}$.
  Although $\mathtt{a}$ diverges as $\sim(r-3/4)^{-2+\sqrt{2}}$ at the universal horizon, this does not result in the divergences of the curvature invariants.}
  \label{figs}
\end{figure}

        \begin{figure}[tb]
                \begin{center}
                                \includegraphics[keepaspectratio=true,height=55mm]{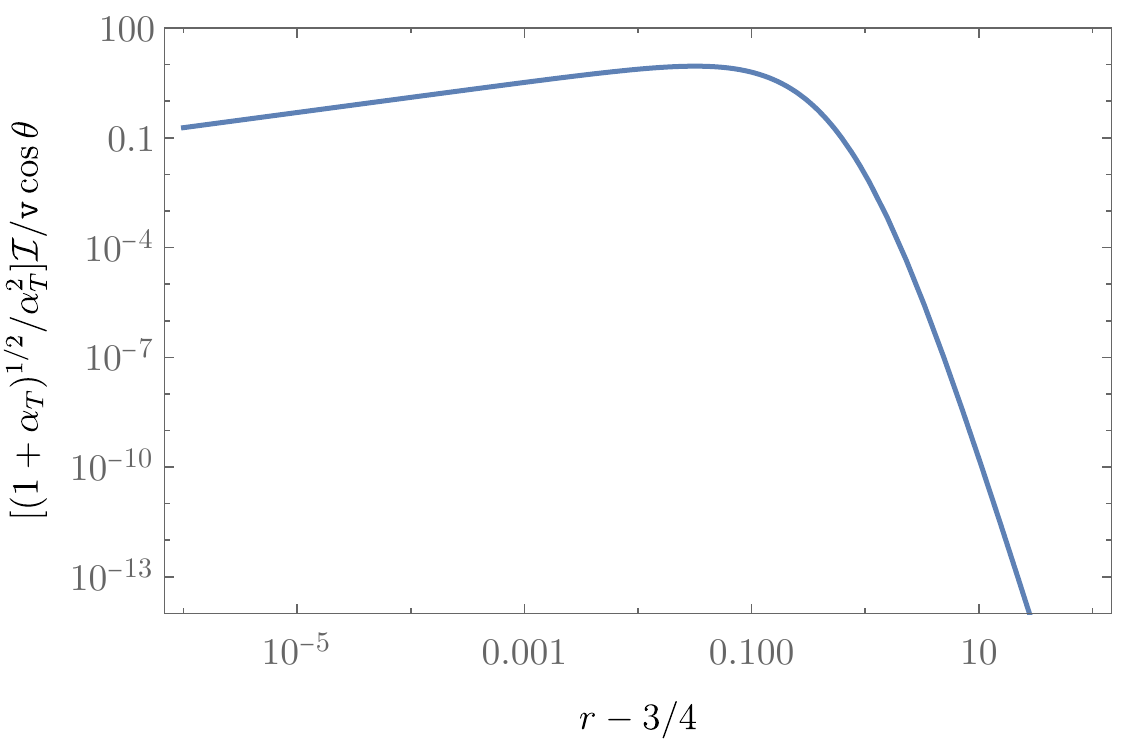}
                \end{center}
                                \caption{Curvature invariant $\mathcal{I}$ as a function of $r$.
                                The $\alpha_T$ dependence is isolated as the overall factor $(1+\alpha_T)^{1/2}/\alpha_T^2$.}
                \label{fig:curv-I.pdf}
        \end{figure}

The final step is to confirm that the metric perturbations thus obtained satisfy the boundary conditions~\eqref {eq:bc-inf} and hence describe a slowly moving black hole.
For large $r$, Eq.~\eqref{eq:def-zeta} (with $\zeta=0$) reduces to 
\begin{align}
    \chi''+\frac{2}{r}\chi'-\frac{2}{r^2}\chi=0,
\end{align}
and the growing solution to this equation is 
\begin{align}
    \chi= \mathcal{C}_1r.
\end{align}
By numerically integrating Eq.~\eqref{eq:def-zeta} with the boundary condtion $\chi=\mathcal{C}_+(r-3/4)^{-1+\sqrt{2}}$ at the universal horizon, we find that $\mathcal{C}_1\simeq 1.00568\times \mathcal{C}_+$.
Substituting this growing solution to Eqs.~\eqref{eq:eta-1}--\eqref{eq:eta-4}, we see that
\begin{align}
    \mathsf{H}_1\approx -\frac{2}{9}\mathcal{C}_1,
    \quad 
    \mathsf{H}_2\approx 0,
    \quad 
    \mathsf{a}\approx 0,
    \quad 
    \mathsf{b}\approx -\frac{2}{9}\mathcal{C}_1r
    \quad (r\to \infty),
\end{align}
which shows that the solution satisfies the boundary conditions~\eqref{eq:bc-inf} with $\mathtt{v}=-2\mathcal{C}_1/9$.
One can thus find a slowly moving black hole solution that is regular outside the universal horizon for any value of $\alpha_T$ and for any nonvanishing function $\alpha_L(X)$.
The reconstructed metric perturbations are presented in Fig.~\ref{figs}.
The curvature invariant $\mathcal{I}$ is shown in Fig.~\ref{fig:curv-I.pdf}.

Our results are consistent with the decoupling limit analysis of Ref.~\cite{Kovachik:2023fos}, where regular khronon configurations are found for any values of the parameters of the theory.
The non-analyticity of the khronon configurations at the universal horizon in~\cite{Kovachik:2023fos} is translated to the divergence of derivatives of $\mathcal{I}$ in our analysis.
In the case of $\alpha_T=0$, there is no such non-analyticity in $\mathcal{R}$ (and its derivatives) because we have simply $\mathcal{R}=0$.
As we do not see any $\alpha_L$-dependent effects in the solution, it is just the Schwarzschild geometry expressed in a nontrivial coordinate system.
Our results in the case of $\alpha_T=0$ therefore not only agree with those of Ref.~\cite{Ramos:2018oku}, but also generalize them to theories with general $\alpha_L(X)$.
However, in the case of $\alpha_T\neq 0$, our results are in contradiction with Ref.~\cite{Ramos:2018oku}, where the linear perturbation contribution to $\mathcal{R}$ seems to diverge as $\sim(r-3/4)^{-2-\sqrt{2}}$ at the universal horizon (as far as we can visually observe in Fig.~3 of Ref.~\cite{Ramos:2018oku}).
Although the origin of this discrepancy is not completely clear, we guess that this divergence comes from the $\mathcal{Z}_-$ mode because $\mathcal{R}\sim \mathcal{Z}_-(r-3/4)^{-2-\sqrt{2}}$ as we have seen above.

\section{Conclusions}\label{sec:concluding}

If the Lorentz symmetry is broken by a timelike gradient of a scalar field, a black hole solution moving with respect to the rest frame of the scalar field is not obtained by a boost from the solution at rest.
In this paper, we have studied such moving black solutions in a subset of higher-order scalar-tensor theories.
The family of theories we have considered is characterized by a constant $\alpha_T$, which parametrizes the deviation of the speed of gravitational waves from that of light, and an arbitrary nonvanishing function $\alpha_L(X)$ of the kinetic term of the scalar field $X$.
A black hole has been assumed to be slowly moving so that it can be described in terms of static, even-parity, dipole perturbations.

Closely related analyses in the context of khronometric theory and Ho\v{r}ava-Lifshitz gravity are found in Refs.~\cite{Ramos:2018oku,Kovachik:2023fos}, and we have generalized and improved these previous studies in several aspects.
First, our results can be applied not only to khronometric theory but also to more general theories such as cuscuton-like gravity with no propagating scalar degree of freedom~\cite{Gao:2019twq}.
Second, we have consistently considered full metric perturbations and a scalar-field perturbation, extending the decoupling limit analysis.
A consistent treatment of the metric and scalar-field perturbations has already been presented in~\cite{Ramos:2018oku}, but we have refined the previous analysis by deriving a single master equation from which all the metric perturbations can be reconstructed.
Our master equation and the metric reconstruction allow one to see clearly how the results depend on $\alpha_T$ and $\alpha_L$.

By requiring that the curvature invariants are finite at the universal horizon, we have shown that a moving black hole solution is completely independent of $\alpha_L(X)$, while the $\alpha_T$ dependence appears as a rather trivial rescaling of the metric components.
In the case of $\alpha_T=0$, this implies that the metric is Schwarzschild expressed in a nontrivial coordinate system, which is in agreement with the previous result derived for $\alpha_L=\,$ const~\cite{Ramos:2018oku}.
However, contrary to~\cite{Ramos:2018oku}, we have found a moving black hole solution for which the curvature invariant is finite everywhere outside the universal horizon in the case of $\alpha_T\neq 0$.
This result is rather consistent with the decoupling limit analysis of~\cite{Kovachik:2023fos}.


As demonstrated in Appendix~\ref{sec:app-pert-Mink}, we have an infinite propagation speed of the scalar mode around Minkowski spacetime, $c_s\to\infty$, implying potential pathologies.
This is not problematic if one specializes to $\alpha_L$ given by Eq.~\eqref{eq:2dof-alpha-L}, because there is no propagating scalar mode around any background from the beginning~\cite{Gao:2019twq}.
In the cases of generic $\alpha_L(X)$, we need, for example, to introduce nonvanishing $\beta_3$ to keep $c_s$ finite and remove the concerns about pathologies, as has been done in the context of Ho\v{r}ava-Lifshitz gravity~\cite{Blas:2009qj}.
Nevertheless, in this paper, we have focused on the family of theories defined by Eq.~\eqref{eq:def-functions}, because the analysis in this paper has been simplified due to the limit $c_s\to \infty$.
It would be interesting to go beyond the $\beta_3\to 0$ limit and extend the present analysis to include the effects of finite $c_s$.
We hope to come back to this issue in a future publication.

\acknowledgments
The work of JS was supported by the Rikkyo University Special Fund for Research.
The work of TK was supported by
JSPS KAKENHI Grant No.~JP20K03936 and
MEXT-JSPS Grant-in-Aid for Transformative Research Areas (A) ``Extreme Universe'',
No.~JP21H05182 and No.~JP21H05189.

\appendix

\section{Perturbations around Minkowski}\label{sec:app-pert-Mink}

Let us study the behavior of the scalar mode around Minkowski spacetime in the theory described by the unitary gauge action~\eqref{eq:ADM-action}.
To do so, we start with the action with one additional term,
\begin{align}
    S=\frac{1}{2}\int\D t\D^3x\sqrt{\gamma}N \left[
        K_{ij}K^{ij}-\left(1+\frac{2}{3}\alpha_L\right)K^2 
        +(1+\alpha_T)R+\beta_3a_ia^i
    \right],\label{eq:action-app-adm}
\end{align}
where $a_i=\partial_i\ln N$.
Here, $\beta_3$ is defined in Eq.~\eqref{def:eft-paras} and may depend on $N$ in the unitary gauge.
Note that $\alpha_L$ and $\alpha_T$ may also depend on $N$, but the $N$-dependence of these coefficients does not matter in the following analysis.
Therefore, the essential results have already been derived in the context of the extension of Ho\v{r}ava-Lifshitz gravity in~\cite{Blas:2009qj}.
After deriving the quadratic action for the scalar mode, we take the limit $\beta_3\to 0$.

The scalar perturbations in the unitary gauge are given by
\begin{align}
    N=1+\delta N(t,\Vec{x}),\quad N_i=\partial_iw(t,\Vec{x}),
    \quad \gamma_{ij}=e^{2\zeta(t,\Vec{x})}\delta_{ij}.\label{scalar_pert:Minkowski}
\end{align}
Substituting these ADM variables to the action~\eqref{eq:action-app-adm}
and expand it to second order in perturbations, we obtain 
\begin{align}
    S=\int\D^4 x\left[
        -3(1+\alpha_L)\dot\zeta^2+2(1+\alpha_L)\dot\zeta\partial^2w-\frac{\alpha_L}{3}
        (\partial^2w)^2-2(1+\alpha_T)\delta N\partial^2\zeta 
        +(1+\alpha_T)(\partial\zeta)^2+\frac{\beta_3}{2}(\partial\delta N)^2
    \right],\label{eq:2nd-action_Min}
\end{align}
The Euler-Lagrange equations for $\delta N$ and $w$ reduce to constraint equations, which can be solved to give 
\begin{align}
    \delta N=-\frac{2}{\beta_3}(1+\alpha_T)\zeta ,
    \quad 
    \partial^2w=\frac{3(1+\alpha_L)}{\alpha_L}\dot \zeta.
\end{align}
Substituting these back to the action, we obtain 
\begin{align}
    S=\int\D t\D^3x \left[
    \frac{3(1+\alpha_L)}{\alpha_L}\dot\zeta^2-\frac{(1+\alpha_T)(2-\beta_3)}{\beta_3}(\partial\zeta)^2
    \right].
\end{align}

We require $A_L=-\alpha_L/(1+\alpha_L)<0$ to avoid the ghost instability.
The propagation speed of the scalar mode is given by 
\begin{align}
    c_s^2=c_{\textrm{GW}}^2\cdot \frac{(-A_L)}{3}\cdot \frac{2-\beta_3}{\beta_3}.
\end{align}
Given that $c_{\textrm{GW}}^2=1+\alpha_T>0$, $c_s^2$ is positive for $0<\beta_3<2$.
In the limit $\beta_3\to 0$, we have $c_s^2\to\infty$, implying that the scalar mode is infinitely strongly coupled.

\section{Static monopole perturbations}\label{app:mono}

Let us present some results on static monopole perturbations of the black hole solution presented in Sec.~\ref{sec:BHsoln}.
With an appropriate gauge fixing, the static monopole perturbations are given by~\cite{Saito:2023bhn}
\begin{align}
        \delta g_{\mu\nu}\D x^\mu\D x^\nu =-\mathsf{H}_0\D t^2+\frac{2\mathsf{H}_1}{\sqrt{f}}\D t\D r
        +\frac{\mathsf{H}_2}{f}\D r^2,
        \quad \delta\phi=0.
\end{align}
We introduce the new variables $\mathsf{h}_0$ and $\mathsf{h}_1$ defined by
\begin{align}
        \mathsf{H}_0&=\left(f-B^2\right)\mathsf{h}_0-2B\mathsf{h}_1-f\mathsf{H}_2, 
        \\ 
        \mathsf{H}_1&=\frac{B}{2}\mathsf{h}_0+\mathsf{h}_1+\frac{B}{2}\mathsf{H}_2,
\end{align}
and replace $\mathsf{H}_0$ and $\mathsf{H}_1$ with them.
Then, the quadratic action for the monopole perturbations is given by
\begin{align}
        S=4\pi\int\D t\D r\mathcal{L},
\end{align}
with 
\begin{align}
        \mathcal{L}=-\frac{\alpha_Lr^2}{12}\left[
                \frac{2}{r^2}\left(r^2\mathsf{h}_1\right)'+B\mathsf{h}_0'
        \right]^2-r\left[B\mathsf{h}_1+\frac{1}{2}(1+\alpha_T)f\mathsf{H}_2
        \right]
        \mathsf{h}_0'.
\end{align}
The Euler-Lagrange equations derived from this read
\begin{align}
        \mathsf{h}_0'&=0,
        \\ 
        \left[\frac{\alpha_L}{r^2}
        \left(r^2\mathsf{h}_1\right)'\right]'&=0,
        \\ 
        \left[
                rB\mathsf{h}_1+\frac{1}{2}(1+\alpha_T)rf\mathsf{H}_2
        \right]'&=0.
\end{align}
We impose the boundary conditions at infinity as $\mathsf{H}_2,\mathsf{h}_0,\mathsf{h}_1\to 0$.
The solution is then obtained as
\begin{align}
        \mathsf{h}_0&=0,
        \\ 
        \mathsf{h}_1&=\frac{(1+\alpha_T)^{1/2}\delta b_0}{r^2}
        +\frac{\mathcal{C}_0}{r^2}\int^r\frac{r^2}{\alpha_L}\D r,
        \\ 
        B\mathsf{h}_1+\frac{1}{2}(1+\alpha_T)f\mathsf{H}_2&=
        \frac{1}{2}(1+\alpha_T)\frac{\delta\mu}{r},
\end{align}
where $\delta\mu$, $\delta b_0$, and $\mathcal{C}_0$ are integration constants.
Notice that $\delta\mu$ and $\delta b_0$ correspond, respectively, to the shifts of the integration constants of the background solution, $\mu$ and $b_0$.
Since $\alpha_L$ is a function of $X$ and $X \to 1/\sqrt{2}N_0$ as $r\to\infty$,
we may assume that $\alpha_L\to$ const as $r\to\infty$.
Therefore, for the boundary condition $\mathsf{h}_1\to 0$ to be satisfied at infinity,
we need to impose $\mathcal{C}_0=0$.
To summarize, we have no nontrivial solutions for static monopole perturbations of the black hole solution studied in the main text.

\section{The explicit expressions for the coefficients in Eqs.~\eqref{eq:eta-1}--\eqref{eq:eta-4}}\label{sec:app-eta-cfs}

The coefficients appearing in the reconstructed metric perturbations~\eqref{eq:eta-1}--\eqref{eq:eta-4} are given explicitly as follows:
\begin{align}
        \eta_{\mathsf{H}_1}^{(1)}&=\frac{3}{32r^6}\frac{f^{1/2}}{f'},
        \\ 
        \eta_{\mathsf{H}_1}^{(2)}&=-\left(\frac{2}{9 r^4}+\frac{81}{4096 r^{10}}\right)
        \frac{f^{3/2}}{(f')^2},
        \\
        \eta_{\mathsf{H}_1}^{(3)}&=\left(\frac{8}{81 r^5}+\frac{5}{81 r^6}+\frac{1}{27 r^7}-\frac{1}{32
        r^8}-\frac{5}{384 r^9}-\frac{1}{512 r^{10}}+\frac{99}{2048
        r^{11}}+\frac{99}{4096 r^{12}}+\frac{297}{32768 r^{13}}\right)
        \frac{(r-3/4)^2}{f^{3/4}(f')^2},
        \\ 
        \tilde \eta_{\mathsf{H}_1}^{(3)}&=\frac{3}{64r^5}\frac{1}{f^{3/4}f'},
        \\  
        \eta_{\mathsf{H}_1}^{(4)}&=-\left(\frac{1}{96 r^3}+\frac{4}{81}\right)\frac{f^{1/4}}{f'},
        \\ 
        \eta_{\mathsf{H}_2}^{(1)}&=
        \frac{1}{r^4}\frac{f^{1/2}}{f'},
        \\ 
        \eta_{\mathsf{H}_2}^{(2)}&=
        -\left(\frac{1}{r^5}+\frac{27}{128 r^8}\right)\frac{f^{3/2}}{(f')^2},
        \\
        \eta_{\mathsf{H}_2}^{(3)}&=
        \left(-\frac{1}{3 r^6}-\frac{1}{12 r^7}+\frac{1}{16 r^8}+\frac{45}{64
        r^9}+\frac{45}{128 r^{10}}+\frac{135}{1024 r^{11}}\right)
        \frac{(r-3/4)^2}{f^{3/4}(f')^2}
        ,
        \\ 
        \tilde \eta_{\mathsf{H}_2}^{(3)}&=\frac{1}{2r^3}\frac{1}{f^{3/4}f'},
        \\  
        \eta_{\mathsf{H}_2}^{(4)}&=-\frac{1}{3r}\frac{f^{1/4}}{f'},
        \\ 
        \eta_{\mathsf{a}}^{(1)}&=\frac{1}{2r^4} 
        \frac{(r-3/4)}{f^{1/2}f'}
        ,
        \\ 
        \eta_{\mathsf{a}}^{(2)}&=\left(-\frac{1}{2 r^5}+\frac{3}{8 r^6}+\frac{9}{32 r^7}+\frac{27}{256
        r^8}\right)
        \frac{(r-3/4)}{f^{1/2}f'}
        ,
        \\ 
        \eta_{\mathsf{a}}^{(3)}&=\left(
                -\frac{1}{6 r^4}-\frac{1}{6 r^5}-\frac{1}{8 r^6}+\frac{3}{16
   r^7}-\frac{9}{256 r^8}-\frac{27}{1024 r^9}-\frac{81}{4096
   r^{10}}
        \right)\frac{(r-3/4)}{f^{3/4}(f')^2},
        \\ 
        \tilde \eta_{\mathsf{a}}^{(3)}&=\frac{1}{4r^2}\frac{1}{f^{3/4}f'},
        \\ 
        \eta_{\mathsf{a}}^{(4)}&=-\frac{1}{6}\frac{1}{f^{3/4}(f')^2},
        \\ 
        \eta_{\mathsf{b}}^{(1)}&=-\frac{2}{9}\left(1-\frac{1}{r}\right)
        ,
        \\ 
        \eta_{\mathsf{b}}^{(2)}&=\frac{3}{64r^4}\frac{f}{f'}
        ,
        \\ 
        \eta_{\mathsf{b}}^{(3)}&=\left(-\frac{7}{96 r^3}-\frac{1}{192 r^4}+\frac{9}{2048 r^7}-\frac{1}{81
        r}+\frac{8}{81}\right)\frac{1}{f^{1/4}f'},
        \\ 
        \eta_{\mathsf{b}}^{(4)}&=-\frac{4 r^3}{81}f^{3/4}.
\end{align}

\bibliography{refs}

\providecommand{\href}[2]{#2}\begingroup\raggedright\begin{thebibliography}{10}

\bibitem{Frolov:2023gsl}
V.P.~Frolov, \emph{{Motion of a rotating black hole in a homogeneous scalar field}}, \href{https://doi.org/10.1103/PhysRevD.109.024055}{\emph{Phys. Rev. D} {\bfseries 109} (2024) 024055} [\href{https://arxiv.org/abs/2312.07801}{{\ttfamily 2312.07801}}].

\bibitem{Frolov:2024htj}
V.P.~Frolov and A.~Koek, \emph{{Motion of rotating black holes in homogeneous scalar fields: A general case}}, \href{https://doi.org/10.1103/PhysRevD.109.084067}{\emph{Phys. Rev. D} {\bfseries 109} (2024) 084067} [\href{https://arxiv.org/abs/2403.01044}{{\ttfamily 2403.01044}}].

\bibitem{Liberati:2013xla}
S.~Liberati, \emph{{Tests of Lorentz invariance: a 2013 update}}, \href{https://doi.org/10.1088/0264-9381/30/13/133001}{\emph{Class. Quant. Grav.} {\bfseries 30} (2013) 133001} [\href{https://arxiv.org/abs/1304.5795}{{\ttfamily 1304.5795}}].

\bibitem{Blas:2014aca}
D.~Blas and E.~Lim, \emph{{Phenomenology of theories of gravity without Lorentz invariance: the preferred frame case}}, \href{https://doi.org/10.1142/S0218271814430093}{\emph{Int. J. Mod. Phys. D} {\bfseries 23} (2015) 1443009} [\href{https://arxiv.org/abs/1412.4828}{{\ttfamily 1412.4828}}].

\bibitem{Herrero-Valea:2023zex}
M.~Herrero-Valea, \emph{{The status of Ho\v{r}ava gravity}}, \href{https://doi.org/10.1140/epjp/s13360-023-04593-y}{\emph{Eur. Phys. J. Plus} {\bfseries 138} (2023) 968} [\href{https://arxiv.org/abs/2307.13039}{{\ttfamily 2307.13039}}].

\bibitem{Horava:2009uw}
P.~Horava, \emph{{Quantum Gravity at a Lifshitz Point}}, \href{https://doi.org/10.1103/PhysRevD.79.084008}{\emph{Phys. Rev. D} {\bfseries 79} (2009) 084008} [\href{https://arxiv.org/abs/0901.3775}{{\ttfamily 0901.3775}}].

\bibitem{Blas:2009qj}
D.~Blas, O.~Pujolas and S.~Sibiryakov, \emph{{Consistent Extension of Horava Gravity}}, \href{https://doi.org/10.1103/PhysRevLett.104.181302}{\emph{Phys. Rev. Lett.} {\bfseries 104} (2010) 181302} [\href{https://arxiv.org/abs/0909.3525}{{\ttfamily 0909.3525}}].

\bibitem{Blas:2010hb}
D.~Blas, O.~Pujolas and S.~Sibiryakov, \emph{{Models of non-relativistic quantum gravity: The Good, the bad and the healthy}}, \href{https://doi.org/10.1007/JHEP04(2011)018}{\emph{JHEP} {\bfseries 04} (2011) 018} [\href{https://arxiv.org/abs/1007.3503}{{\ttfamily 1007.3503}}].

\bibitem{Horava:2010zj}
P.~Horava and C.M.~Melby-Thompson, \emph{{General Covariance in Quantum Gravity at a Lifshitz Point}}, \href{https://doi.org/10.1103/PhysRevD.82.064027}{\emph{Phys. Rev. D} {\bfseries 82} (2010) 064027} [\href{https://arxiv.org/abs/1007.2410}{{\ttfamily 1007.2410}}].

\bibitem{Blas:2009yd}
D.~Blas, O.~Pujolas and S.~Sibiryakov, \emph{{On the Extra Mode and Inconsistency of Horava Gravity}}, \href{https://doi.org/10.1088/1126-6708/2009/10/029}{\emph{JHEP} {\bfseries 10} (2009) 029} [\href{https://arxiv.org/abs/0906.3046}{{\ttfamily 0906.3046}}].

\bibitem{Jacobson:2010mx}
T.~Jacobson, \emph{{Extended Horava gravity and Einstein-aether theory}}, \href{https://doi.org/10.1103/PhysRevD.81.101502}{\emph{Phys. Rev. D} {\bfseries 81} (2010) 101502} [\href{https://arxiv.org/abs/1001.4823}{{\ttfamily 1001.4823}}].

\bibitem{Jacobson:2000xp}
T.~Jacobson and D.~Mattingly, \emph{{Gravity with a dynamical preferred frame}}, \href{https://doi.org/10.1103/PhysRevD.64.024028}{\emph{Phys. Rev. D} {\bfseries 64} (2001) 024028} [\href{https://arxiv.org/abs/gr-qc/0007031}{{\ttfamily gr-qc/0007031}}].

\bibitem{Jacobson:2007veq}
T.~Jacobson, \emph{{Einstein-aether gravity: A Status report}}, \href{https://doi.org/10.22323/1.043.0020}{\emph{PoS} {\bfseries QG-PH} (2007) 020} [\href{https://arxiv.org/abs/0801.1547}{{\ttfamily 0801.1547}}].

\bibitem{Ramos:2018oku}
O.~Ramos and E.~Barausse, \emph{{Constraints on Ho\v{r}ava gravity from binary black hole observations}}, \href{https://doi.org/10.1103/PhysRevD.99.024034}{\emph{Phys. Rev. D} {\bfseries 99} (2019) 024034} [\href{https://arxiv.org/abs/1811.07786}{{\ttfamily 1811.07786}}].

\bibitem{Kovachik:2023fos}
A.~Kovachik and S.~Sibiryakov, \emph{{Slowly moving black holes in khrono-metric model}},  \href{https://arxiv.org/abs/2311.12936}{{\ttfamily 2311.12936}}.

\bibitem{Langlois:2015cwa}
D.~Langlois and K.~Noui, \emph{{Degenerate higher derivative theories beyond Horndeski: evading the Ostrogradski instability}}, \href{https://doi.org/10.1088/1475-7516/2016/02/034}{\emph{JCAP} {\bfseries 02} (2016) 034} [\href{https://arxiv.org/abs/1510.06930}{{\ttfamily 1510.06930}}].

\bibitem{DeFelice:2018ewo}
A.~De~Felice, D.~Langlois, S.~Mukohyama, K.~Noui and A.~Wang, \emph{{Generalized instantaneous modes in higher-order scalar-tensor theories}}, \href{https://doi.org/10.1103/PhysRevD.98.084024}{\emph{Phys. Rev. D} {\bfseries 98} (2018) 084024} [\href{https://arxiv.org/abs/1803.06241}{{\ttfamily 1803.06241}}].

\bibitem{Gao:2019twq}
X.~Gao and Z.-B.~Yao, \emph{{Spatially covariant gravity theories with two tensorial degrees of freedom: the formalism}}, \href{https://doi.org/10.1103/PhysRevD.101.064018}{\emph{Phys. Rev. D} {\bfseries 101} (2020) 064018} [\href{https://arxiv.org/abs/1910.13995}{{\ttfamily 1910.13995}}].

\bibitem{Afshordi:2006ad}
N.~Afshordi, D.J.H.~Chung and G.~Geshnizjani, \emph{{Cuscuton: A Causal Field Theory with an Infinite Speed of Sound}}, \href{https://doi.org/10.1103/PhysRevD.75.083513}{\emph{Phys. Rev. D} {\bfseries 75} (2007) 083513} [\href{https://arxiv.org/abs/hep-th/0609150}{{\ttfamily hep-th/0609150}}].

\bibitem{Langlois:2017mxy}
D.~Langlois, M.~Mancarella, K.~Noui and F.~Vernizzi, \emph{{Effective Description of Higher-Order Scalar-Tensor Theories}}, \href{https://doi.org/10.1088/1475-7516/2017/05/033}{\emph{JCAP} {\bfseries 05} (2017) 033} [\href{https://arxiv.org/abs/1703.03797}{{\ttfamily 1703.03797}}].

\bibitem{Horndeski:1974wa}
G.W.~Horndeski, \emph{{Second-order scalar-tensor field equations in a four-dimensional space}}, \href{https://doi.org/10.1007/BF01807638}{\emph{Int. J. Theor. Phys.} {\bfseries 10} (1974) 363}.

\bibitem{Deffayet:2011gz}
C.~Deffayet, X.~Gao, D.A.~Steer and G.~Zahariade, \emph{{From k-essence to generalised Galileons}}, \href{https://doi.org/10.1103/PhysRevD.84.064039}{\emph{Phys. Rev. D} {\bfseries 84} (2011) 064039} [\href{https://arxiv.org/abs/1103.3260}{{\ttfamily 1103.3260}}].

\bibitem{Kobayashi:2011nu}
T.~Kobayashi, M.~Yamaguchi and J.~Yokoyama, \emph{{Generalized G-inflation: Inflation with the most general second-order field equations}}, \href{https://doi.org/10.1143/PTP.126.511}{\emph{Prog. Theor. Phys.} {\bfseries 126} (2011) 511} [\href{https://arxiv.org/abs/1105.5723}{{\ttfamily 1105.5723}}].

\bibitem{Crisostomi:2016czh}
M.~Crisostomi, K.~Koyama and G.~Tasinato, \emph{{Extended Scalar-Tensor Theories of Gravity}}, \href{https://doi.org/10.1088/1475-7516/2016/04/044}{\emph{JCAP} {\bfseries 04} (2016) 044} [\href{https://arxiv.org/abs/1602.03119}{{\ttfamily 1602.03119}}].

\bibitem{Langlois:2018dxi}
D.~Langlois, \emph{{Dark energy and modified gravity in degenerate higher-order scalar\textendash{}tensor (DHOST) theories: A review}}, \href{https://doi.org/10.1142/S0218271819420069}{\emph{Int. J. Mod. Phys. D} {\bfseries 28} (2019) 1942006} [\href{https://arxiv.org/abs/1811.06271}{{\ttfamily 1811.06271}}].

\bibitem{Kobayashi:2019hrl}
T.~Kobayashi, \emph{{Horndeski theory and beyond: a review}}, \href{https://doi.org/10.1088/1361-6633/ab2429}{\emph{Rept. Prog. Phys.} {\bfseries 82} (2019) 086901} [\href{https://arxiv.org/abs/1901.07183}{{\ttfamily 1901.07183}}].

\bibitem{Saito:2024xdx}
J.~Saito, Z.~Yao and T.~Kobayashi, \emph{{PPN meets EFT of dark energy: post-Newtonian approximation in higher-order scalar-tensor theories}}, \href{https://doi.org/10.1088/1475-7516/2024/06/040}{\emph{JCAP} {\bfseries 06} (2024) 040} [\href{https://arxiv.org/abs/2402.10459}{{\ttfamily 2402.10459}}].

\bibitem{Will:2018bme}
C.M.~Will, \emph{{Theory and Experiment in Gravitational Physics}}, Cambridge University Press (9, 2018).

\bibitem{Li:2009bg}
M.~Li and Y.~Pang, \emph{{A Trouble with Ho\v{r}ava-Lifshitz Gravity}}, \href{https://doi.org/10.1088/1126-6708/2009/08/015}{\emph{JHEP} {\bfseries 08} (2009) 015} [\href{https://arxiv.org/abs/0905.2751}{{\ttfamily 0905.2751}}].

\bibitem{Henneaux:2009zb}
M.~Henneaux, A.~Kleinschmidt and G.~Lucena~G\'omez, \emph{{A dynamical inconsistency of Horava gravity}}, \href{https://doi.org/10.1103/PhysRevD.81.064002}{\emph{Phys. Rev. D} {\bfseries 81} (2010) 064002} [\href{https://arxiv.org/abs/0912.0399}{{\ttfamily 0912.0399}}].

\bibitem{Iyonaga:2018vnu}
A.~Iyonaga, K.~Takahashi and T.~Kobayashi, \emph{{Extended Cuscuton: Formulation}}, \href{https://doi.org/10.1088/1475-7516/2018/12/002}{\emph{JCAP} {\bfseries 12} (2018) 002} [\href{https://arxiv.org/abs/1809.10935}{{\ttfamily 1809.10935}}].

\bibitem{Berglund:2012bu}
P.~Berglund, J.~Bhattacharyya and D.~Mattingly, \emph{{Mechanics of universal horizons}}, \href{https://doi.org/10.1103/PhysRevD.85.124019}{\emph{Phys. Rev. D} {\bfseries 85} (2012) 124019} [\href{https://arxiv.org/abs/1202.4497}{{\ttfamily 1202.4497}}].

\bibitem{Barausse:2011pu}
E.~Barausse, T.~Jacobson and T.P.~Sotiriou, \emph{{Black holes in Einstein-aether and Horava-Lifshitz gravity}}, \href{https://doi.org/10.1103/PhysRevD.83.124043}{\emph{Phys. Rev. D} {\bfseries 83} (2011) 124043} [\href{https://arxiv.org/abs/1104.2889}{{\ttfamily 1104.2889}}].

\bibitem{Blas:2011ni}
D.~Blas and S.~Sibiryakov, \emph{{Horava gravity versus thermodynamics: The Black hole case}}, \href{https://doi.org/10.1103/PhysRevD.84.124043}{\emph{Phys. Rev. D} {\bfseries 84} (2011) 124043} [\href{https://arxiv.org/abs/1110.2195}{{\ttfamily 1110.2195}}].

\bibitem{Saito:2023bhn}
J.~Saito and T.~Kobayashi, \emph{{Black hole perturbations in spatially covariant gravity with just two tensorial degrees of freedom}}, \href{https://doi.org/10.1103/PhysRevD.108.104063}{\emph{Phys. Rev. D} {\bfseries 108} (2023) 104063} [\href{https://arxiv.org/abs/2308.00267}{{\ttfamily 2308.00267}}].

\bibitem{Lin:2017jvc}
K.~Lin, S.~Mukohyama, A.~Wang and T.~Zhu, \emph{{No static black hole hairs in gravitational theories with broken Lorentz invariance}}, \href{https://doi.org/10.1103/PhysRevD.95.124053}{\emph{Phys. Rev. D} {\bfseries 95} (2017) 124053} [\href{https://arxiv.org/abs/1704.02990}{{\ttfamily 1704.02990}}].

\end{thebibliography}\endgroup
\bibliographystyle{JHEP}
\end{document}